\documentclass[12pt]{article}
\usepackage{epsfig}
\newcommand{\sig}[1]{$e \quad \mu \,\tau \, \tau \, b\:$}

\begin{document}
\title{ \bf \large Multilepton Signatures of the Higgs Boson through its Production in Association
with a Top-quark Pair}
\author{Pankaj Agrawal, Somnath Bandyopadhyay and Siba Prasad Das\footnote{email:
agrawal@iopb.res.in, somnath@iopb.res.in and spdas@iopb.res.in}\\
Institute of Physics \\ Sachivalaya Marg, Bhubaneswar, Orissa, India 751 005}
\maketitle
\begin{abstract}

We consider the possible production of the Higgs Boson in association with a 
top-quark pair and its subsequent decay into a tau-lepton pair or a W-boson pair. 
This process can 
give rise to many signatures of the Higgs boson. These signatures can have 
electrons, muons, tau jets, bottom jets and/or light flavour jets. 
We analyze the viability of some of these signatures. We will look at those 
signatures where the background is minimal. In particular, we explore the viability of
the signatures ``isolated 4 electron/muon'' and ``isolated 3 electron/muon + a jet'' 
The jet can be due to a light flavour quark/gluon, a bottom quark, or a tau lepton. 
Of all these signatures, we find that ``isolated 3 electron/muon + a tau jet'',
with an extra bottom jet, can be an excellent signature of this mode of the Higgs 
boson production. We show that this signature may be visible within a year,
once the Large Hadron Collider (LHC) restarts. Some of the other signatures would 
also be observable after the LHC accumulates sufficient luminosity.
\end{abstract}

\newpage

\section{Introduction}

     The Standard Model (SM) has been enormously successful \cite{discoverh,Chatrchyan:2012ufa}. 
     Until recently,  one important ingredient of the model, the Higgs mechanism, 
     had no direct experimental support. The implementation of the Higgs mechanism 
     through  a set of scalar fields has been a standard paradigm, 
     which is also used in a variety of the extensions of the Standard Model (SM) 
     to break the gauge symmetries and bring it to the level of the SM.  
     One consequence of the Higgs mechanism is the existence of the 
     scalar particles. The number and nature of the particles depend on 
     the symmetry that has been broken. 

       In the SM, the mechanism gives rise to a neutral scalar particle -- 
     the Higgs boson. In the run I (2009-12) of the Large Hadron Collider (LHC), 
     strong evidence for a Higgs-boson like particle has been found \cite{discoverh}.
     Some of its properties like spin and mass have also been measured by
     the CMS \cite{Chatrchyan:2012ufa} and ATLAS collaborations \cite{Aad:2012mea}. 
     Combining the signal of the Higgs boson from its various decay modes, more
     than $5 \sigma$ enhancement above the background is seen by both
     collaborations. It has all but confirmed the existence of the Higgs 
     boson. Its mass is expected to be around 125 GeV.

     One of the main goals of the run II of the LHC (2015-18) would be 
     to establish the existence of the Higgs boson more firmly and really 
     show that its is a SM Higgs boson scalar.
     To show that this particle is indeed a SM particle,
     and does not belong to its extensions or modifications, it would be 
     important to identify the Higgs boson through multiple production
     mechanisms and decay channels. There are many important production 
     mechanisms, like gluon fusion, W-fusion, associated production with a vector
     boson and the production in association with a bottom-quark pair 
     or  top-quark pair \cite{hsmlhc}. For
     a 125 GeV Higgs boson, there are a number of important decay channels --
     $ H \to b {\bar b} \cite{pahiggsbb}, W W^{*}$ \cite{atlashwwsm,Kao:hww}, $Z Z^{*} \; {\rm and} \; \tau \tau$ 
     \cite{Chatrchyan:2011nx,Chatrchyan:2012vp,atlasHtautau,taupibonn,Baglio:2011xz,pahiggstautau}. All these
     major production and decay channels (including rare decays like 
     $H \to \gamma \gamma$ \cite{atlastthgammagamma,cmsgamma}) will be observable in the run II of the LHC.
     Some of these channels have already been seen in the run I \cite{hsmlhc}.

     In this letter, we focus on the production mechanism $p p \to t {\bar t} H$, at 
     $\sqrt s$= 14 TeV, with the subsequent decay of the Higgs boson into a 
     tau-lepton pair \cite{cmstautau}, or a W-boson pair.
     There are enormous possibilities for a variety of signatures because
     there are many heavy particles in the final state which then decay
     into many more particles. In this letter, we will look at those
     signatures which have most leptons in the final state.  More leptons
     in the final state means smaller background. However, it
     comes at the cost of fewer signal events. In a subsequent paper \cite{abd}, we
     will analyze the signatures which have fewer leptons and more jets. There
     we will have more signal events, but larger background.

     In the next section, we will discuss production, decay and signatures
     in a bit more detail. In the section 3, we would discuss the backgrounds.
     In the section 4, we would present numerical results. In the last section,
     we would conclude.

\section{Production, Decay and Signatures}

      We are considering the production of the Higgs boson with a top-quark
      pair. This is fourth most important production mechanism. The
      process occurs through gluon-gluon or quark-quark annihilation.
      We will consider semileptonic decay of both the top quarks and the
      decay of the Higgs boson into a tau-lepton pair, or a W-boson pair. 
      For the $M_H = 120 - 130\,$ GeV, the tau-lepton decay mode has a 
      branching ratios of a $5 - 7$ percent. The tau-lepton
      can further decay into an electron/muon or hadrons and neutrinos. 
      When it decays into hadrons, it manifest itself as a jet -- tau jet. This jet
      has special characteristics compared to a quark/gluon jet. It
      is narrow and has very few hadrons. It is narrow because of the
      low mass of the tau lepton; it has few hadrons because tau lepton
      mostly has one-prong or three-prong decays. These properties of
      a tau jet can be used to distinguish it from a quark/gluon jet.
      The W-boson decay mode of the Higgs boson has a branching ratio
      of $14-30 \%$ for the Higgs boson with the mass in the $120-130\,$ GeV
      range. Here both W-boson cannot be on-shell. The W-boson decays into
      leptons/quarks and neutrinos.

      This production and decay chain can give rise to a multitude of 
      signatures. The final state can have only jets, one electron/muon
      and jets, two electron/muon and jets, three electron/muon and jets
      and four electron/muon and jets. Some of these jets can be bottom
      jets or/and tau jets. Of all these signatures, because of the
      larger branching ratios, ``only jets'' signature will give rise
      to most signal events; but it will also have the largest background
      due to the production of the jets through the strong interaction
      processes.  On the other hand, we have a signature of ``4 electron/muon
      + jets''. This signature has least number of signal events, but
      also the smallest background. One of the drawback of all these
      signatures is that one cannot reconstruct the Higgs boson mass
      through its decay products. This is because of the presence
      of many neutrinos in its decay products. However, as we will see,
      due to the manageable background, we can still identify the
      Higgs boson through these production and decay chains.

       We shall consider the signature of `` 4 electron/muon
      + jets'' and ``3 electron/muon + jets''. Because of
      the small cross section for such events, due to small semileptonic
      branching ratios, we would minimize the number of jets to be 
      observed. This will help us in increasing the number of signal events 
      marginally, without increasing the background. So in the end, we
      shall be considering four signatures: ``4 electron/muon'',
      ``3 electron/muon + a jet'', ``3 electron/muon + a tau jet'',
      and ``3 electron/muon + a bottom-jet''. In this list,
       ``3 electron/muon + a jet'' will have largest signal events,
       while ``4 electron/muon'' will have the least number of
       signal events. We will also consider the signature ``3 electron/muon'' alone.
       The numerical results would be presented for three of these signatures,
       as the other two have large backgrounds.

       Let us first consider the signature: ``4 electron/muon''.
       Such events occur when both the top quarks and tau leptons decay
       semileptonically. Such events also receive contributions when
       the Higgs boson (in $t {\bar t} H$ production) decays into 2 W-bosons. We
       will see that it makes larger contribution. Another contribution 
       comes from the process $ gg \to H$ and $H \to Z Z^{*}$. Such a
       contribution will be reduced if we veto events with a lepton pair
       of same flavor opposite charge (SFOC) which has mass close to the 
       mass of the Z-boson. We have not included these events in
       the signature. Other signatures, 
       with 3 electron/muon, occur when out of the top-quark pair and the 
       tau-lepton pair, only three particles decay semileptonically;
       the remaining particle decays into hadrons/tau jet. These 
       events also receive contribution from the decay $H \to W W^{*}$
       after the $ttH$ production. In this case, tau-lepton decay mode
       makes larger contribution. Because of the decay of the top
       quarks, these events naturally have bottom jets, irrespective of
       whether we observe them or not. We will find that observation
       of an extra bottom jet can increase the significance of a signature.
       We can also have a real tau jet
       in the signal events through the Higgs boson or a top-quark decay.
      
\section{Backgrounds}

       All the signatures under consideration will receive contribution
       from the signal events, i.e. the production of the Higgs boson, and
       other SM processes which does not have a Higgs boson.
       Question is -- is the background small enough to be sure that signal
       events have been produced ? To establish the viability of the
       signatures, we shall first identify the major background processes
       and then estimate their contributions.
       There are two classes of the backgrounds: (1)
       direct backgrounds, (2) mimic backgrounds. In the case of 
       the direct background, the background processes produce events
       similar to the signal events. They have same particles as in
       the signal. On the other hand, mimic backgrounds
       have jets, which can mimic (fake) a tau jet, a bottom jet, or
       even an electron/muon. These mimic probabilities are usually 
       quite small -- less than a percent. So even if a background
       has large cross section, it becomes smaller when folded
       with mimic probability.

       \begin{enumerate}
       \item ``4 electron/muon'': There are many 
       processes which can be backgrounds. The source of direct
       backgrounds are the processes $t {\bar t} Z, WWZ, WWWW, ZZ, t {\bar t}t {\bar t}$.
       The main sources of mimic backgrounds are:  $WZ + {\rm jet}, 
       t {\bar t} W, WWW + {\rm jet}$. These background occur when a 
       jet mimics an electron/muon. As discussed below, the mimic backgrounds are 
       not significant because of the very small probability of a jet to mimic an
       electron/muon, about $10^{-5}$ \cite{jetlep}.

         Among the direct backgrounds, the most significant backgrounds would be due
       to the production of $ t {\bar t} Z$ and $ZZ$ events and subsequent decay
       into leptons. Using {\tt MadGraph v5} \cite{Maltoni:2002qb}, we find that 
       the cross sections for the signal
       $t {\bar t} H$ is about 0.44 pb for $m_H = 125$ GeV, while the cross sections for
        $ t {\bar t} Z$ and $ZZ$ are 0.66 pb and 10.8 pb respectively. Because of very
       similar structure,  $ t {\bar t} Z$ will always be a significant background
       to the signal. These two backgrounds can be reduced by requiring
       appropriate $M_{\ell_1 \ell_2}$ to be away from the mass of the Z-boson. But the
       background when a Z-boson decays into a tau-lepton pair and subsequent decay
       of the tau-leptons into electron/muon cannot be reduced in this way. These and the 
       other values of the cross sections from MadGraph are with its default settings,
       unless stated otherwise. The processes $WWZ, WWWW, {\rm and} \,\, t {\bar t}t {\bar t} $ 
       have the cross sections of about 100.0, 0.6 and 12.0 fb respectively. We clearly 
       see that these processes are not important source of
       the backgrounds due to small cross sections.

       \item ``3 electron/muon + a jet'': In this case,
       the direct backgrounds are $t {\bar t} Z,  t {\bar t} W, ZZ,  WZ + {\rm jet},
       WWW + {\rm jet}, WWZ$; the major mimic backgrounds are $t {\bar t}$ and
       $WW +  2 {\rm jet}$. As above, due to small probability of a jet faking an 
       electron/muon, the mimic backgrounds can be ignored. Most of the direct backgrounds
       are self-explanatory. $ZZ$ production is a background, when a Z-boson decays
       into a tau-lepton pair, and one of the tau leptons appears as a tau jet.

       \item ``3 electron/muon + a tau jet'': In this case,
       the direct backgrounds are $t {\bar t} Z,  t {\bar t}t {\bar t} , WWZ$;
       the major mimic backgrounds are $t {\bar t}, WZ + {\rm jet},
       WW +  2 \;\;{\rm jet},  WWW + {\rm jet}, t {\bar t} W$.  As above, the
       backgrounds that fake an lepton are not important. But that backgrounds
       $ WZ + {\rm jet}$ and $t {\bar t} W$ can be important where a light/bottom
       jet mimics a tau jet.

       \item ``3 electron/muon + a bottom jet'': In this case,
       the direct backgrounds are $t {\bar t} Z, t {\bar t} W, t {\bar t}t {\bar t} $.
       In  these processes the bottom jet would appear from a top-quark decay.
       The major mimic backgrounds are $t {\bar t}, WZ + {\rm jet},
       WW +  2 \;\; {\rm jet},  WWZ, WWW + {\rm jet}, t {\bar t} W$. 

       \item ``3 electron/muon'': There are many 
       processes which can be backgrounds. The source of direct
       backgrounds are the processes $t {\bar t} Z, t {\bar t} W,
       WWZ, WWW, WZ, t {\bar t}t {\bar t} $.
       The main sources of mimic backgrounds are: $WW + {\rm jet}, 
       t {\bar t}$. These background occur when a jet mimics an electron/muon.

       \end{enumerate}

\section{Numerical results and Discussion}

         In this section, we are presenting numerical results. The signal and
	 the background calculations have been done using {\tt ALPGEN v2.14} \cite{alpgen} 
         and its interface with {\tt PYTHIA v6.325}\cite{pythia}. Using {\tt ALPGEN},
         we generate the parton-level unweighted events.
	 These events are then turned into more realistic events by hadronization,
	 initial and final state radiation using {\tt PYTHIA}. We 
         have also applied  following generic kinematic cuts:
  
$$
p_T^{e, \mu,j} > 20 \; \rm{GeV},\: |\eta^{e, \mu,j}| < 2.5,\; R(jj,\ell j, \ell\ell) > 0.4.
$$
         We have used  CTEQ5L \cite{cteq5} parton distribution functions 
         and other default parameters including renormalization and factorization scales. 
         For the results, we have chosen the center-of-mass energy of 14 TeV and integrated
         luminosity is $100 \;$ fb$^{-1}$. We take mass of the top quark is 174.3 GeV. 
         We are taking three different values for the mass
         of the Higgs boson -- 120, 125 and 130 GeV.

         One of our signatures has a tau jet. Both CMS and ATLAS collaborations \cite{atlastautau} 
         can identify tau jets. A tau jet is a manifestation of the hadronic decays of a tau lepton.
         A tau lepton has a branching ratios of approximately $65\%$ to decay into hadrons.
         Two main characteristics of a tau jet are its narrowness and presence of only a few
         hadrons. These two features have been used to identify a tau jet. However, like the
         identification of a bottom jet, the identification of tau jet can only be done with
         some probability. The other jets due to quarks/gluon can also mimic a tau jet with
         small probability. Usually there is a trade-off between higher detection efficiency
         and higher rejection of the mimic-jets. We are taking two cases - one with high tau jet
         detection efficiency, other with low tau-detection efficiency. We are also presenting
         results by identifying a tau jet with an area-variable. This variable alone would
         not work well, as our number shows. The number of charge tracks will play a crucial
         role in discriminating a tau jet.

          We are considering the detection of a bottom-jet also. We have used the identification
         probability ($\epsilon_{b}$) of $55\%$\cite{cmsepsb,mdadspd}. For other jets to mimic a bottom jet, 
         we  use the probability
         of $1\%$. For a jet faking a lepton, the probability
         is quite low. A light flavour jet can mimic a lepton with a probability of about $10^{-5}$.
         For a bottom jet such a number is $5 \times 10^{-5}$. As we see in the signature, leptons comes
         either from the decay of a top-quark, or the decay of a tau-lepton, or a W-boson. So a
         pair of leptons would not have mass near the mass of a Z-boson. But a number of backgrounds
         have a Z-boson, so we use a cut of invariant mass of SFOC leptons: 
         $|M_{\ell_1 \ell_2} - M_Z| < 15 \;\; {\rm GeV}$ to reduce these backgrounds.

\begin{figure}[ht!]
\begin{center}
 \raisebox{0.0cm}{\hbox{\includegraphics[angle=-90,scale=0.33]{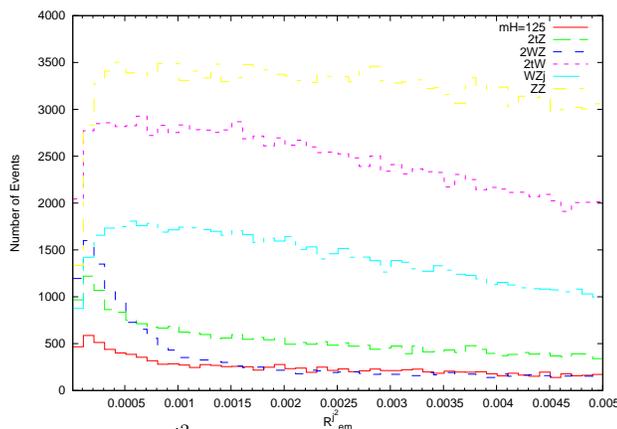}}}
\caption{ The profile of $R^{j^2}_{em}$  for the signal and major SM backgrounds.} 
\label{r2jem}
\end{center}
\end{figure}

          We are presenting the results for the three signatures: ``3 electron/muon + a tau jet'',
         ``3 electron/muon + a bottom jet'', and ``4 electron/muon''. In Table 1,
          we present the results for ``3 electron/muon + a tau jet''. For the
	  signal events, the largest contribution comes from the tau-lepton decay channel of the 
	  Higgs boson. The Contribution of this channel is about $75\%$. The contribution of the
	  W-boson decay channel is about $25\%$. Here we have considered four
	  cases. In the first case, R-cut, we have used an area-variable of the tau jet cone, 
          $R^{j^2}_{em}$ (adapted from \cite{Englert:2011iz}) to identify a tau jet. 
          The behavior of the variable for the signal and backgrounds without normalization are 
          displayed in Fig.~\ref{r2jem}. We clearly see that in the processes where there is  
          a tau jet the variable is peaked towards a low value.
          We have checked that the area-variable gives better tau-jet efficiencies than 
	  the radius ($R^{j}_{em}$) \cite{abd}. We have used a cut of $R^{j^2}_{em} < 1 \times 10^{-4}$. 
          As we are using only one characteristic of the tau jet,
	  its narrowness, so it is not necessarily the best way \cite{dutta}.
          We have tau-identification rate of
	  $30\%$ and mimic (rejection) rate of about $3\%$. In the second case of LTT, low tau-tagging,
	  we have taken the low value for the tau-jet identification, $27\%$, and low mimic
	  rate of $0.25\%$. Compared to the case 1, the signal decreases a bit and some of the backgrounds,
	  specially $WZ + jet$ and $t \bar{t} W$ reduce significantly. Therefore, the significance of the
	  signature increases. The case 3 of HTT\cite{cmsepstau}, high tau-tagging, has high identification rate of
	  $50\%$ and the mimic rate of $1\%$. We see that the significance of the signature increases
	  again. This is because of the larger number of signal events. In the case 4, we have used
	  the fact that some of the backgrounds do not have a bottom jet for ``free''. So if we observe
	  an extra bottom jet, i. e., the signature `` 3 electron/muon + a tau jet + a bottom jet'',
	  then the background will reduce further, thus enhancing the significance of the signature.
	  Since there are two bottom jets and only one is to be identified, we have used the 
	  identification probability of $80\%$.
          We will note that without identification  of a jet some of the backgrounds would be higher
	  by two-orders of magnitude, making the signal harder to observe. So identification of a
	  jet play important role in reducing the backgrounds.

	  In Table 2, we present the results for ``3 electron/muon + a bottom-jet''. So
	  we wish to identify a bottom jet instead of a tau jet. We now have fewer major backgrounds.
	  But the $t {\bar t} W$ background increases by more than a order of magnitude. This is
	  because this process has a bottom jet, and there is no need for this jet to mimic a tau jet.
          Therefore this is not an attractive signature, but with enough integrated luminosity, this
	  signature can be observed.

	  In Table 3, there are results for the ``4 electron/muon'' signature. In this case,
	  $75\%$ of the events are through the W-boson decay channel of the Higgs boson; the
	  rest are from the tau-lepton decay channel.  We notice
	  that this is an observable signature with a significance 3 -- 5, depending on the
	  mass of the Higgs boson. This signature is also obtained by the $g g \to H
	  \to Z Z ^{*}$ process \cite{atlasfourlep}.
          So we also look for an extra bottom jet to make the signature
	  exclusive for the $t {\bar t} H$ process. The major background is $t {\bar t} Z$
	  process. We see that the  signature `` 4 electron/muon + a bottom-jet'' is a useful signature
	  with significance approaching 5 with 100 fb$^{-1}$ of integrated luminosity. 

\vskip 0.5in

\begin{tabular}{||c|c|c|c|c|c|c|c|c|c|c|c||} \hline
 & \multicolumn{3}{c|}{Signal, $M_H$ (GeV)} & \multicolumn{5}{c|}{Backgrounds} & \multicolumn{3}{c||}{$S/\sqrt{B}$, $M_H$ (GeV)} \\ \cline{2-12}
  $\tau$ jets id & 120  & 125 & 130 & $t {\bar t} Z$ & $W W Z$ & $t {\bar t} W$ & $W Z$j & $ZZ$ & 120  & 125 & 130 \\ \hline
R-cut   &  22  &  20 & 19 & 14 & 2 & 12 & 24 & 7 & 2.9 & 2.6 & 2.5\\
LTT   &  20  & 18  &  17  & 13 & 2 & 1 & 2 & 6 & 4.1 & 3.7 & 3.5 \\
HTT  &   37 &  33  &  32  & 23 & 3 & 4 & 10 & 12 & 5.1 & 4.6 & 4.4 \\
B-tag/HTT  & 30 & 27  & 26 & 18 & 0 & 3  & 0 & 0 & 6.6 & 5.9& 5.7\\ \hline
\end{tabular}
\vskip .3in
{\small
Table 1: Number of events for the signature ``3 electron/muon + tau jet''
at the LHC with the integrated luminosity of 100 fb$^{-1}$ with
the cuts and efficiencies specified in the text.}

\vskip 0.5in
\begin{center}

\begin{tabular}{||c|c|c|c|c|c|c|c|c||} \hline
 \multicolumn{3}{||c|}{Signal, $M_H$ (GeV)} & \multicolumn{3}{c|}{Backgrounds} & \multicolumn{3}{c||}{$S/\sqrt{B}$, $M_H$ (GeV)}\\ \hline
  120  & 125 & 130 & $t {\bar t} Z$ & $t {\bar t} W$ & $W Z$j & 120 & 125 & 130\\ \hline
 42   & 34   &  26  & 26 & 312 &  6 & 2.3 & 1.8 & 1.4\\ \hline
\end{tabular}
\vskip .3in
{\small
Table 2: Number of events for the signature ``3 electron/muon + bottom jet''
at the LHC with the integrated luminosity of 100 fb$^{-1}$ with
the cuts and efficiencies specified in the text.}

\end{center}
\vskip 0.5in

\begin{center}

\begin{tabular}{||c|c|c|c|c|c|c|c|c|c||} \hline
  & \multicolumn{3}{c|}{Signal, $M_H$ (GeV)} & \multicolumn{3}{c|}{Backgrounds}  & \multicolumn{3}{c||}{$S/\sqrt{B}$, $M_H$ (GeV)}\\ \cline{2-10}
  bottom jet id & 120  & 125 & 130 & $t {\bar t} Z$ & $WWZ$ & ZZ & 120 & 125 & 130 \\ \hline
no extra b   &  16  & 19  &  22  & 15 & 2 & 3 & 3.1 & 4.3 & 4.9  \\
extra b  &  13  & 16   & 18   & 12  & 0 & 0  & 3.8 & 4.6 & 5.2 \\ \hline
\end{tabular}
\vskip .3in
{\small
Table 3: Number of events for the signature ``4 electron/muon''
at the LHC with the integrated luminosity of 100 fb$^{-1}$ with
the cuts and efficiencies specified in the text.}

\end{center}

\vskip 0.2in

\section{Conclusion}

         In this letter, we have considered ``4 electron/muon'' and ``3 
         electron/muon + jet'' signatures of the process $p p \to t {\bar t} H$. Here jet can be
         a light flavour quark/gluon  jet, a tau jet, or a bottom jet. We find that, of all
         these signatures, ``3 electron/muon + a tau jet'', specially with an extra
	 bottom jet observation, i.e., ``3 electron/muon + a tau jet + a bottom jet'',
	 appears to be the most promising signature. With 100 fb$^{-1}$ of luminosity, 
	 it has the significance
	 of 5.9 for $M_H = 125$ GeV. This signature may be observable in 
	 about a year of running of the LHC in run II. The signature  
	 ``4 electron/muon +  bottom jet'' is a distinctive signature
	 of the $p p \to t {\bar t} H$ process and it should also be observable
	 within a year of run II.
	 The signatures  ``3 electron/muon + a bottom jet'' 
         and  ``4 electron/muon'' should also be observable in the run II. A more
         detailed analysis of these and other signatures of the Higgs boson, when it is produced
         in association with a pair of top-quarks, will be presented elsewhere.

\end{document}